\documentclass{aa}

\usepackage{graphicx}
\usepackage{txfonts}

\usepackage{amsmath}    
\usepackage{amssymb}    
\usepackage{mathtools}
\usepackage{multicol} 
\usepackage{multirow} 
\usepackage{threeparttable}
\usepackage{longtable}
\usepackage{adjustbox}

\usepackage{color}
\usepackage{natbib,twoopt}
\usepackage[hyphenbreaks]{breakurl}
\usepackage[breaklinks]{hyperref}      
\bibpunct{(}{)}{;}{a}{}{,}             
\definecolor{cobalt}{rgb}{0.06, 0.2, 0.65}
\hypersetup{
  colorlinks,
  citecolor=blue,
  linkcolor=blue,
  urlcolor=blue,
}
\makeatletter
  \newcommandtwoopt{\citeads}[3][][]{\href{http://adsabs.harvard.edu/abs/#3}%
    {\def\hyper@linkstart##1##2{}%
     \let\hyper@linkend\@empty\citealp[#1][#2]{#3}}}
  \newcommandtwoopt{\citepads}[3][][]{\href{http://adsabs.harvard.edu/abs/#3}%
    {\def\hyper@linkstart##1##2{}%
     \let\hyper@linkend\@empty\citep[#1][#2]{#3}}}
  \newcommandtwoopt{\citetads}[3][][]{\href{http://adsabs.harvard.edu/abs/#3}%
    {\def\hyper@linkstart##1##2{}%
     \let\hyper@linkend\@empty\citet[#1][#2]{#3}}}
  \newcommandtwoopt{\citeyearads}[3][][]%
    {\href{http://adsabs.harvard.edu/abs/#3}
    {\def\hyper@linkstart##1##2{}%
     \let\hyper@linkend\@empty\citeyear[#1][#2]{#3}}}
\makeatother

\newcommand{\Msun}{M$_{\odot}$}
\newcommand{\Rsun}{R$_{\odot}$}

\definecolor{smalt(darkpowderblue)}{rgb}{0.0, 0.2, 0.6}
\definecolor{forestgreen(traditional)}{rgb}{0.0, 0.5, 0.0}

\defcitealias{RVJ}{RVJ}

\defcitealias{MD17}{MD17}

\newcommand{\afgk}{AFGK-type}

\newcommand{\bse}{{\sc bse}}

\begin{document}

   \title{Formation of long-period post-common envelope binaries}

   \subtitle{I. No extra energy is needed to explain oxygen-neon white dwarfs paired with AFGK-type main-sequence stars}

   \titlerunning{Formation of long-period PCEBs I. No extra energy needed for massive WD plus MS}

   \author{Diogo Belloni\inst{1}
           \and
           Monica Zorotovic\inst{2}
           \and
           Matthias R. Schreiber\inst{1,3}
           \and
           Steven G. Parsons\inst{4}
           \and
           Maxwell Moe\inst{5}
           \and
           James A. Garbutt\inst{4}
          }

    \authorrunning{Belloni et al.}

   \institute{Departamento de F\'isica, Universidad T\'ecnica Federico Santa Mar\'ia, Av. España 1680, Valpara\'iso, Chile\\
              \email{diogobellonizorzi@gmail.com}
              \and
              Instituto de F\'isica y Astronom\'ia, Universidad de Valpara\'iso, Av. Gran Breta\~na 1111, Valpara\'iso, Chile
              \and
              Millenium Nucleus for Planet Formation, Valpara{\'i}so, Chile
              \and
              Department of Physics and Astronomy, University of Sheffield, Sheffield, S3 7RH, UK
              \and
              Department of Physics \& Astronomy, University of Wyoming, Laramie, WY 82071
             }

   \date{Received...; accepted ...}

 
  \abstract
   {
   It has been claimed for more than a decade that energies other than orbital and thermodynamic internal are required to explain post-common envelope (CE) binaries with sufficiently long orbital periods (${\gtrsim1}$~d) hosting AFGK-type main-sequence stars (${\sim0.5-2.0}$~\Msun) paired with oxygen-neon white dwarfs (${\gtrsim1.1}$~\Msun). This would imply a completely different energy budget during CE evolution for these post-CE binaries in comparison to the remaining systems hosting M dwarfs and/or less massive white dwarfs.
   }
   {
   In this first in a series of papers related to long-period post-CE binaries, we investigated whether extra energy is required  to explain the currently known post-CE binaries with sufficiently long orbital periods consisting of oxygen-neon white dwarfs with AFGK-type main-sequence star companions.
   }
   {
   We carried out binary population simulations with the BSE code adopting empirically derived inter-correlated main-sequence binary distributions for the initial binary population and assuming that the only energy, in addition to orbital, that  help to unbind the CE is   thermal energy. We also searched for the formation pathways of the currently known systems from the zero-age main-sequence binary to their present-day observed properties.
   }
   {
   Unlike what has been claimed for a long time, we show that all such post-CE binaries can be explained by assuming inefficient CE evolution, which is consistent with results achieved for the remaining post-CE binaries. There is therefore no need for an extra energy source. We also found that for CE efficiency close to 100\%, post-CE binaries hosting oxygen-neon white dwarfs with orbital periods as long as one thousand days can be explained. For all known systems we found formation pathways consisting of CE evolution triggered when a highly evolved (i.e. when the envelope mass is comparable to the core mass), thermally pulsing, asymptotic giant branch star fills its Roche lobe at an orbital period of several thousand days. Due to the sufficiently low envelope mass and sufficiently long orbital period, the resulting post-CE orbital period can easily be several tens of days.
   }
   {
   We conclude that the known post-CE binaries with oxygen-neon white dwarfs and AFGK-type main-sequence stars can be explained without invoking any energy source other than orbital and thermal energy. Our results strengthen the idea that the most common formation pathway of the overall population of post-CE binaries hosting white dwarfs is through inefficient CE evolution. 
   }


   \maketitle
%


\section{Introduction}
\label{introduction}

Most close binaries containing stellar remnants are believed to form during common envelope (CE) evolution \citep[e.g.][]{Paczynski_1976,Ivanova_REVIEW,BelloniSchreiberChapter}, in which friction drastically reduces the orbital separation and part of the orbital energy is used to unbind the CE, leaving the exposed core of the giant and its companion in a much tighter orbit.
Post-CE binaries are thus close binaries that survived the engulfment of a star by the deep convective envelope of a companion that filled its Roche lobe when it was a red giant.

In most cases (i.e. for initial masses below ${\sim8-10}$~\Msun) the core of the red giant   cools down to become a white dwarf (WD).
If the Roche lobe was filled during the first ascent on the red giant branch (FGB) it will be a low-mass (${\lesssim0.5}$~\Msun) helium core WD or it will be  a low-mass (${\sim0.32-0.47}$~\Msun) hybrid carbon-oxygen core WD (as helium contributes   a non-negligible fraction to their masses) if it experienced a phase of core helium burning (as a helium star) after losing the envelope \citep[e.g.][]{Heber_2009,Heber_2016}.
On the other hand, if the Roche lobe was filled during the asymptotic giant branch (AGB), the resulting WD will be more massive (${\gtrsim0.5}$~\Msun) and typically composed of carbon-oxygen. 
If the progenitor star was massive enough to reach a core mass of ${\gtrsim1.1}$~\Msun, the carbon in the core can be converted to oxygen and neon, and consequently the emerging WD would be made of oxygen and neon.

Companions to WDs in observed post-CE binaries can be  unevolved stars (i.e. main-sequence stars), stellar remnants (i.e. helium stars, WDs, neutron stars, or black holes), red giants or subgiants (that most likely were on the main sequence during CE evolution and evolved after the post-CE binary was already formed), brown dwarfs \citep{Zorotovic_2022}, or even planets \citep[e.g.][]{Lagos_2021}.
Throughout this paper, we considered only post-CE binaries consisting of white dwarfs with main-sequence star companions that did not experience mass transfer, that is, an episode of mass transfer in addition to CE evolution in which the main-sequence stars are donors.
Such systems are most suitable for constraining CE evolution because their stellar and orbital parameters have been shaped by CE evolution alone.

Previous attempts to reconstruct the CE phase for observed post-CE binaries have found that the vast majority of systems, with typical orbital periods of hours to a few days, can be explained assuming inefficient CE evolution \citep[e.g.][]{Zorotovic_2010,Toonen_2013,Camacho_2014,Cojocaru_2017,Belloni_2019,Ge_2022,hernandezetal22-1, Zorotovic_2022,schrebak+fuller23-1,Ge_2024}.
However, there are currently several systems with orbital periods that range  from a few tens to a few hundreds of days. These periods are much longer than those of most post-CE binaries \citep{nebot-gomez-moranetal11-1}, but too short to be the result of dynamically stable mass transfer.
It has been claimed that these long-period post-CE binaries cannot be explained by CE evolution without contributions from additional energy sources in the CE energy balance \citep[e.g.][]{Davis_2010,Zorotovic_2010,Zorotovic_2014,Yamaguchi_2024}.
To solve this issue, it has been suggested that a small fraction of the available hydrogen recombination energy decreases the binding energy of the CE.
However, it remains a topic of intense discussion whether recombination energy can have any impact on the CE ejection or not \citep[e.g.][]{Soker_2003,Webbink_2008,Ivanova_REVIEW,Nandez_2015,Ivanova_2015,Ivanova_2018,Soker_2018,Grichener_2018,BelloniSchreiberChapter,Ropke_2023}.

In this paper we  investigate whether CE evolution generated when the WD progenitor was a thermally pulsing AGB (TP-AGB) star can explain the characteristics of the observed long-period systems.
We considered all observationally characterized binaries with sufficiently long orbital periods (${>1}$~d), massive WDs (${>1.1}$~\Msun), main-sequence stars of types earlier than M (${>0.5}$~\Msun), and low-eccentricity orbits (${<0.1}$).
In particular, we carried out binary population models to investigate whether extra energies are required  to explain their properties.
We found that extra energy is not required to explain post-CE binaries with orbital periods as long as one thousand days.
Instead, we present reasonable formation pathways for all considered post-CE binaries assuming inefficient CE evolution.

\begin{table*}[!t]
\centering
\caption{Properties of the known long-period binaries with either confirmed or candidate high-mass WDs and AFGK-type main-sequence stars, ordered according to their orbital periods. 
Systems marked with an asterisk have high eccentricities,  most likely descend from initial triple systems, and therefore do not provide constraints on CE evolution.}
\label{TableOBS}
\begin{threeparttable}
\noindent
\setlength\tabcolsep{9pt} 
\renewcommand{\arraystretch}{1.5} 
\begin{tabular}{lrrrrcccc}
\hline
\vspace{-0.1cm}
\multirow{2}{*}{System} & 
\multicolumn{1}{c}{orbital period} & 
\multicolumn{1}{c}{WD mass} & 
\multicolumn{1}{c}{companion mass} & 
\multirow{2}{*}{eccentricity} & 
\multirow{2}{*}{Reference} \\
 & 
 \multicolumn{1}{c}{(d)} & 
 \multicolumn{1}{c}{(\Msun)} &
 \multicolumn{1}{c}{(\Msun)} & 
  & 
  \\
\hline
\multicolumn{6}{c}{\textit{low-eccentricity systems}} \\ 
J2117$+$0332             &   $17.9239\pm0.0001$   & $>1.244\pm0.027$  & $1.11\pm0.03$ & $0.0007\pm0.0002$ &  (4) \\
IK~Peg                   &   $21.7217\pm0.0001$   &  $1.13-1.24$    & $\sim1.7$     & $\sim0$ & (1,2,3) \\
J1111$+$5515             &   $32.1494\pm0.0022$   & $>1.367\pm0.028$  & $1.15\pm0.02$ & $0.0217\pm0.0003$ &  (4) \\
J1314$+$3818             &   $45.5150\pm0.0047$   &  $1.324\pm0.037$  & $0.71\pm0.01$ & $0.0503\pm0.0003$ &  (4) \\
J2034$-$5037             &   $46.1147\pm0.0006$   & $>1.418\pm0.033$  & $0.96\pm0.02$ & $0.0079\pm0.0002$ &  (4) \\
J0107$-$2827             &   $49.0063\pm0.0008$   & $>1.271\pm0.030$  & $0.97\pm0.03$ & $0.0901\pm0.0005$ &  (4) \\
%
%
\multicolumn{6}{c}{\textit{high-eccentricity systems}} \\ 
%
TYC~5451$-$469$-$1$^{*}$ &  $203.5757\pm0.1247$   & $>0.938\pm0.248$  & $1.04\pm0.47$ & $0.2045\pm0.0044$ &  (5) \\
J0353$+$2900$^{*}$       & $1484.344\pm134.116$ &  $1.117\pm0.164$  & $1.95\pm0.13$ & $0.3452\pm0.0410$ &  (5) \\
\hline
\end{tabular}
\begin{minipage}{\linewidth}
\vspace{0.15cm}
References: (1) \citet{Landsman_1993}, (2) \citet{Vennes_1998}, (3) \citet{Lucy_1971}, (4) \citet{Yamaguchi_2024},\\ (5) \citet{Garbutt_2024}
\end{minipage}
\end{threeparttable}
\end{table*}

\section{Observational sample}
\label{Sample}

The properties of the known binaries with long orbital periods (${\gtrsim1}$~d), oxygen-neon WDs (${\gtrsim1.1}$~\Msun) and \afgk~main-sequence stars are presented in Table~\ref{TableOBS}.
In what follows, we provide a brief discussion of the characteristics of these systems 
and identify the post-CE binaries among them.

\subsection{Low-eccentricity systems}
\label{lowecc}

IK~Peg (also known as HR~8210, HD~204188, or J~2126$+$193) was the first member of this class of post-CE binaries.
It hosts a main-sequence star of spectral type A8 (${\sim1.7}$~\Msun). IK\,Peg was discovered as a single-lined spectroscopic binary almost a century ago by \citet{Harper_1928}, who derived an orbital period of $21.724$~d and a low eccentricity (${e=0.027}$).
A reanalysis of the same data by \citet{Lucy_1971} favoured a perfectly circular orbit.
Combining Harper's data with more recent observations, \citet{Vennes_1998} refined the orbital period measurement ($21.72168\pm0.00009$~d) and confirmed the orbit to be circular.

The WD nature of the unseen companion (named EUVE~J2126$+$193) was established by \citet{Wonnacott_1993} and \citet{Landsman_1993}.
While \citet{Wonnacott_1993} derived a WD mass of ${0.985\pm0.03}$~\Msun, \citet{Landsman_1993} estimated a higher mass of ${1.15^{+0.05}_{-0.15}}$~\Msun~by fitting the extreme ultraviolet spectrum with a method strongly dependent on the distance to IK~Peg, which was poorly constrained at that time.
In agreement with these values, \citet{Vennes_1998} determined a range of values for the mass of the WD of ${1.00-1.32}$~\Msun, which can be further constrained to ${1.13-1.24}$~\Msun~taking into account the 
 \textit{Hipparcos} distance of ${42-50}$~pc \citep{Hipparcos_1997}.

\citet{Yamaguchi_2024} recently presented five post-CE binaries containing massive WD candidates (${\gtrsim1.2}$~\Msun) and main-sequence stars with spectral types earlier than M (${\gtrsim0.7}$~\Msun) with long orbital periods (${20-50}$~d).
These systems were discovered as part of a broader search for compact object binaries from the \textit{Gaia} DR3 non-single star catalogue, and considerably extended the sample of such binaries.

\subsection{High-eccentricity systems}
\label{highecc}

\citet{Garbutt_2024} recently cross-matched several catalogues of ultraviolet excess AFGK-type stars with the Gaia DR3 non-single star catalogue to identify more than 200 candidate WDs plus AFGK-type star binaries.
Among the systems they discovered are four long-period (${\sim200-1500}$~d) and high-eccentricity (${\gtrsim0.2}$) systems consisting of a massive WD with a G-type companion star (${\gtrsim0.8}$~\Msun). Two of the companions are unevolved main-sequence stars, which are flagged with an asterisk in Table~\ref{TableOBS}.
We argue in what follows that these binaries most likely originate from triple evolution, instead of binary evolution.

We can rule out CE evolution as the formation channel of these binaries because of the high eccentricities. 
Shortly before and during CE evolution, tidal forces are supposed to be strong enough to circularize the orbit.

If we exclude CE evolution, the remaining binary star formation channel is dynamically stable mass transfer. 
Dynamically stable mass transfer could in principle explain their eccentricities as several mechanisms have been proposed that are in principle able to pump up the eccentricity during stable mass transfer \citep[e.g.][]{BelloniSchreiberChapter}.
However, mass transfer cannot have been dynamically stable in these cases and the reason is twofold.
If the WD progenitor filled its Roche lobe on the early AGB, 
the fact that the final outcome of the mass transfer phase 
is an oxygen-neon WD, implies that their progenitors must have been at least as massive as ${\sim6}$~\Msun~at the beginning of the AGB phase. 
Given that the companions of the WDs are \afgk~main-sequence stars, with masses lower than ${\sim2}$~\Msun, 
the mass ratio when the WD progenitor was on the early AGB star was at least ${\sim3}$, which is much higher than the maximum mass ratio allowing for dynamically stable mass transfer \citep[e.g.][]{Ge_2010,Ge_2015,Ge_2020,Temmink_2023,Henneco_2024}

The only remaining alternative is that these high-eccentricity systems result from triple evolution.
In this picture, either the high-eccentricity systems  descend from systems that were initially triples and the WD is formed due to a merger event of the inner binary merger, or triple dynamics generated by the mass loss of the WD progenitor cause the inner binary separation to decrease.

In the first case, for example, the inner binary might   consist  of stars with initial masses of ${\sim4}$~\Msun~(i.e. a mass ratio close to unity) and an orbital period of ${\sim100}$~d, and the tertiary could be an \afgk~main-sequence star in an eccentric orbit and with an orbital period of several hundred days.
This tertiary could be more or less unaffected by the evolution of the inner binary if the most massive star in the inner binary  fills its Roche lobe when both stars are on the early AGB phase. In this case  the outcome of CE evolution would be a double helium star binary and these helium stars will eventually merge or they will become white dwarfs that eventually merge due to emission of gravitational waves.

Alternatively, the mass loss induced Kozai-Lidov effect \citep{shappee+thompson13-1} can increase the eccentricity and subsequently the orbital separation of the inner binary (due to tidal interactions) in hierarchical triples.
In this case, one would expect the observed eccentric binaries to still be the inner binaries of hierarchical triple systems. 
Interestingly, TYC~5451$-$469$-$1 shows an additional acceleration term in Gaia DR3 that is unlikely due to the spectroscopic period, indicating that this system is probably a triple \citep{Garbutt_2024}.

We would like to mention that a formation channel involving triple evolution is most likely the case for other types of systems that look like post-CE binaries but have high eccentricities, such as   sdB\_b1 \citep{Lei_2023}, which is a helium star paired with a low-mass main-sequence star.
Investigating in more detail the formation pathways for these highly eccentric systems would require performing dedicated simulations for triple stars to understand whether and under what conditions these binaries could emerge from triple star evolution. 
This is far beyond the scope of the present work.
Therefore, these high-eccentricity systems are not considered in what follows.

\section{Binary population models}
\label{BSE}

We carried out post-CE binary population synthesis with the BSE code \citep{Hurley_2002} adopting the methodology described in detail in \citet{Belloni_2024a}.
Briefly, we selected ${\approx6.15\times10^5}$ zero-age main-sequence binaries and picked the primary mass from the canonical \citet{Kroupa_2001} initial mass function in the range between $1$ and $8$~\Msun. 
We used the correlated distributions derived by \citet{MD17}, in which the orbital period distribution depends critically on the primary mass and the binary fraction and the eccentricity and mass-ratio distributions depend on both orbital period and primary mass.
These fitted correlated distributions are the most realistic ones currently available and should be incorporated into binary population models.

For this paper the treatment of CE evolution in \bse~is crucial.
This phase is usually approximated by a simple equation introduced in the 1980s \citep{webbink84-1,livio+soker98} in which the binding energy of the envelope of the red giant donor ($E_\mathrm{bind}$) at the onset of the CE evolution is assumed to be equal to the change in orbital energy during the spiral-in phase ($\Delta E_\mathrm{orb}$) scaled with a parameter $\alpha$, which corresponds to the fraction of the change in orbital energy that is used to unbind the envelope.
We adopted the relation put forward by \citet{Iben_Livio_1993},

\begin{equation}
E_{\rm bind} \ = \ 
\alpha \ \Delta E_{\rm orb} \ = \ - \ 
\alpha \
 \left( \, \frac{G\,M_{\rm d,c}\,M_{\rm a}}{2\,a_f}  \ - \ 
           \frac{G\,M_{\rm d,c}\,M_{\rm a}}{2\,a_i} \, \right) \ ,
\label{EQALPHACE}
\end{equation}

\noindent
where $G$ is the gravitational constant, $E_{\rm bind}$ is the donor envelope binding energy, $E_{\rm orb}$ is the orbital energy, $M_{\rm a}$ is the accretor mass, $M_{\rm d,c}$ is the core mass of the donor, $a_i$ is the semi-major axis at the onset of the CE evolution, and $a_f$ is the semi-major  axis after CE ejection.
We note that this formalism is different from the one adopted by \citet{webbink84-1} and \citet{deKool_1990}.

The binding energy is usually approximated by

\begin{equation}
E_{\rm bind} \ = \ - \ \frac{G\,M_{\rm d}\,(M_{\rm d}\,-\,M_{\rm d,c})}{\lambda \, R_{\rm d}} \ ,
\label{EQLAMBDA}
\end{equation}

\noindent
where $M_{\rm d}$ is the donor mass, $R_{\rm d}$ is the donor radius, and $\lambda$ is the envelope-structure parameter, which depends on the structure of the donor \citep{Dewi_2000,Xu_2010,Loveridge_2011,Klencki_2021,Marchant_2021}.
While some authors use a constant value of $\lambda$ (typically $0.5$ or $1.0$), others calculate it based on the binding energy of the envelope and structure of the star.
However, these calculations are plagued by uncertainties related to the energies that should be considered to calculate the binding energy.

In our calculations, we adopted different efficiencies ($\alpha$ ranging from $0.1$ to $0.9$, in steps of $0.1$).
The envelope-structure parameter $\lambda$ was calculated according to a similar fitting scheme   provided by \citet[][their Appendix A]{Claeys_2014}, which is based on the detailed numerical stellar evolution calculations by \citet{Dewi_2000} and takes into account the structure and the evolutionary stage of the red giant donor and the envelope thermal energy as constrained by the virial theorem (i.e. increasing $\lambda$ by a factor of two).
However, we also checked the influence of our scheme to compute $\lambda$ (Sect.~\ref{BSEvsLambda}). 
Regardless of the scheme we adopted to compute $\lambda$, we always assumed no contributions from energy sources other than gravitational and thermal.

We assumed solar metallicity (${Z=0.02}$) and a constant star formation rate \citep[e.g.][]{Weidner_2004,Kroupa_2013,Recchi_2015,Schulz_2015} over the age of the Galactic disc \citep[${\approx10}$~Gyr,][]{Kilic_2017}.
Unless clearly mentioned, we used the standard \bse~values for all other stellar and binary evolution parameters \citep[e.g.][]{Hurley_2002,Belloni_2018b,Banerjee_2020}.

After a post-CE binary is formed, it evolves towards shorter orbital periods through orbital angular momentum loss.
We included magnetic braking and emission of gravitational waves as mechanisms to remove orbital angular momentum, as described in \citet[][section 2.4, equation 48]{Hurley_2002}. 
Regarding magnetic braking, we adopted the saturated and disrupted magnetic braking prescription with the scaling factors inferred by \citet{Belloni_2024a}.

\begin{figure*}
  \begin{center}
    \includegraphics[width=0.99\linewidth]{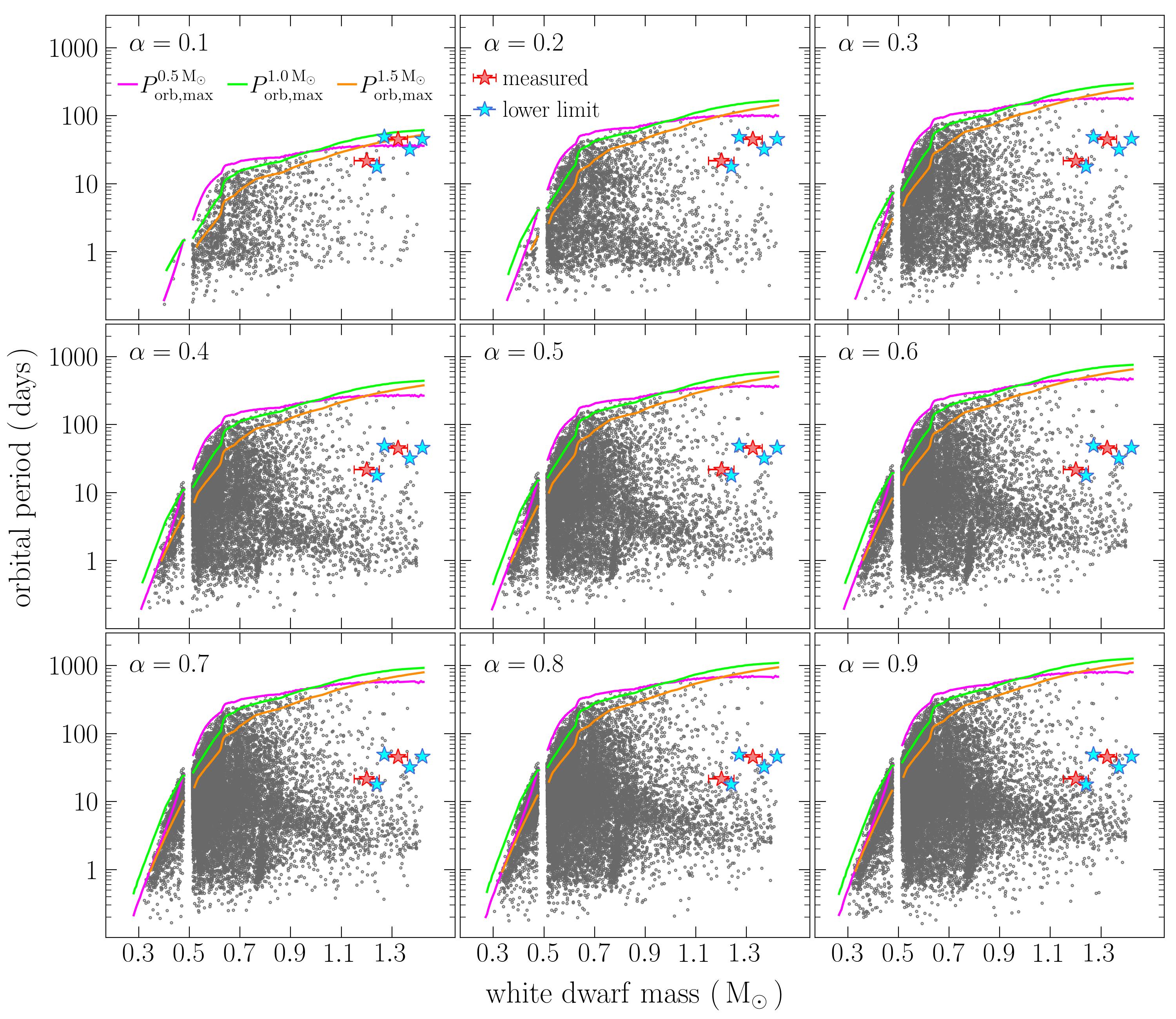}
  \end{center}
  \caption{Distribution of present-day detached post-CE binaries with main-sequence stars of spectral type earlier than M (i.e. masses greater than $0.5$~\Msun) in the plane WD mass vs  orbital period (grey circles), each panel corresponding to a different choice of the CE efficiency $\alpha$. Also included in each panel is the maximum orbital period as a function of the WD mass, for zero-age companion masses of $0.5$~\Msun~(magenta line), $1.0$~\Msun~(green line), and $1.5$~\Msun~(orange line). Also included are the six known long-period post-CE binaries with oxygen-neon WDs and low eccentricity listed in Table~\ref{TableOBS}: two   have measured WD masses (red stars), and five   only have  lower limits for the WD mass (blue stars). In all simulations, no extra energy source other than the orbital, the gravitational, and the thermal was adopted in the CE energy budget. All known systems can be easily explained assuming ${\alpha\gtrsim0.2}$. In addition, for a given WD mass, the greater the  $\alpha$, the longer the maximum orbital period. For ${\alpha\sim1}$, even post-CE binaries with orbital periods as long as a thousand days can be explained.}
  \label{Fig:MwdvsPorb}
\end{figure*}

\begin{figure*}
  \begin{center}
    \includegraphics[width=0.99\linewidth]{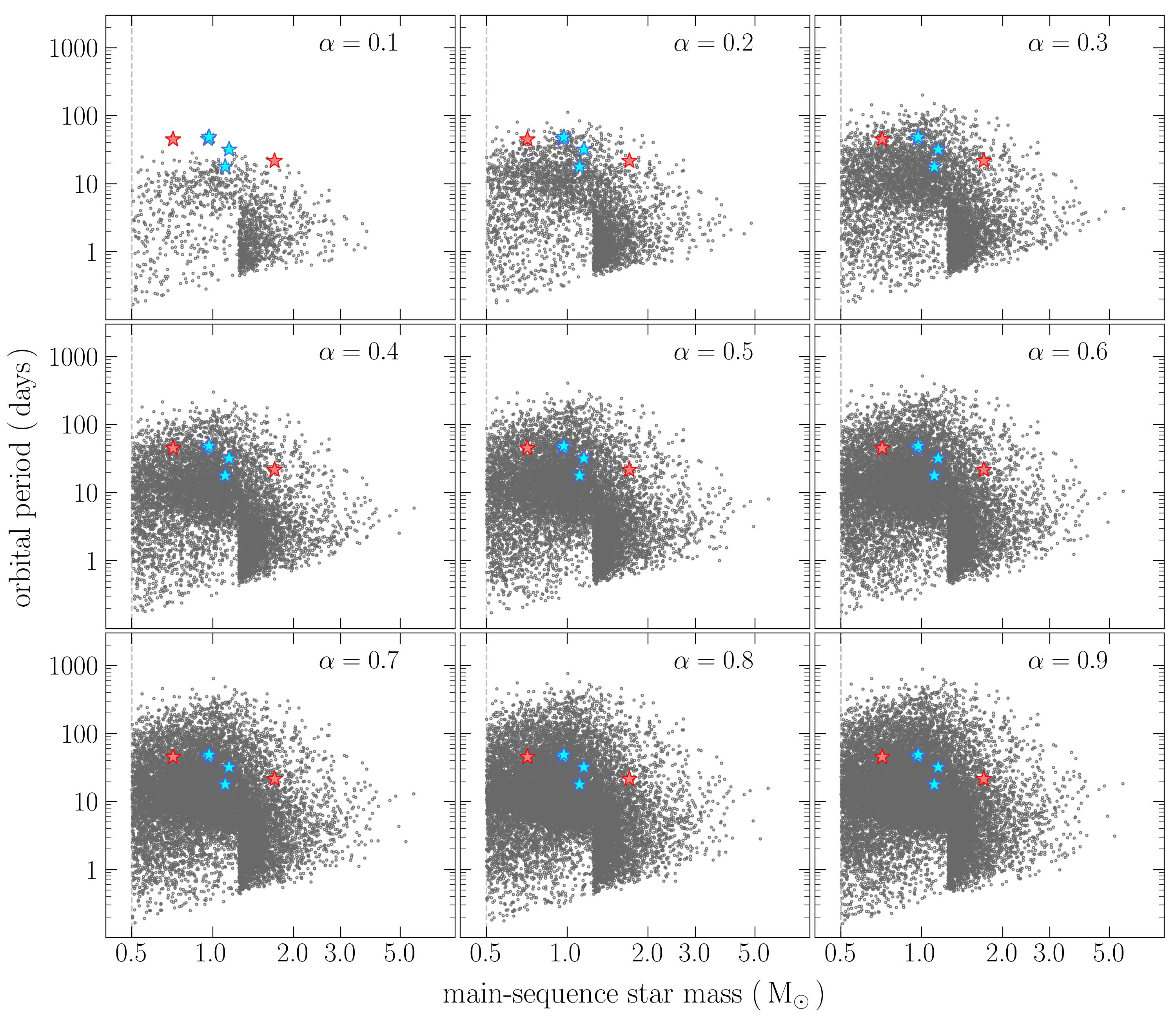}
  \end{center}
  \caption{Similar to Fig.~\ref{Fig:MwdvsPorb}, but in the plane main-sequence companion mass vs orbital period. The detached post-CE binaries are mainly found in two clumps in this plane. The first clump corresponds to main-sequence star masses ${\gtrsim1.25}$~\Msun, which happens because magnetic braking is not expected to be acting in these stars, which means their orbital periods cannot easily change as the only source of orbital angular momentum loss is the emission of gravitational waves. Only those binaries with sufficiently short orbital periods have their orbital periods substantially changed. The second clump corresponds to post-CE binaries with sufficiently long orbital periods (${\gtrsim5}$~d). Irrespective of whether magnetic braking is acting or not in these binaries, their long orbital periods make any source of orbital angular momentum loss weak. For the remaining post-CE binaries (i.e. with main-sequence star masses ${\lesssim1.25}$~\Msun~and orbital periods ${\lesssim5}$~d), magnetic braking combined with emission of gravitational waves more easily turn them into cataclysmic variables, that is, semi-detached binaries, and for this reason this region is less populated. Most importantly, all six known long-period post-CE binaries with massive WDs can be explained for any value of the CE efficiency ${\alpha\gtrsim0.2}$.}
  \label{Fig:M2vsPorb}
\end{figure*}

\section{Results}
\label{Results}

\subsection{Properties of post-CE binaries}
\label{PCEBprop}

We start the presentation of our results by discussing how the CE efficiency $\alpha$ affects the properties of the post-CE binaries.
We show in Figs.~\ref{Fig:MwdvsPorb} and \ref{Fig:M2vsPorb} the main present-day properties (i.e. orbital period, WD mass, and companion mass) of the detached post-CE binaries hosting \afgk~main-sequence stars in our population synthesis, as a function of $\alpha$.
We also included in Fig.~\ref{Fig:MwdvsPorb} the maximum post-CE orbital period for specific zero-age companion masses, namely $0.5$, $1.0$, and $1.5$~\Msun.
During binary evolution, in the case when CE evolution occurs when the more massive star fills its Roche lobe while still on the FGB, the mass of the companion is virtually unaffected as very little of the wind leaving the WD progenitor can be  accreted.
However, if the WD progenitor fills its Roche lobe during TP-AGB evolution, then its companion is able to accrete a substantial amount of the slow wind from the giant prior to CE evolution.
This means that each of the lines in Fig.~\ref{Fig:MwdvsPorb} does not correspond to a fixed companion mass since it can significantly increase depending on the binary evolution before the onset of the CE evolution.

We start our presentation of the predicted post-CE binary populations by inspecting the WD mass versus orbital period plane (Fig.~\ref{Fig:MwdvsPorb}), which shows a clear correlation with $\alpha$.
As expected, the higher the efficiency, the longer the maximum post-CE orbital period, because the orbital energy is used more efficiently to unbind the CE for higher values of $\alpha$, resulting in less orbital shrinkage.

For extremely low-efficiency CE evolution (${\alpha\lesssim0.1}$), post-CE orbital periods are shorter than ${\sim50}$~days.
The lower the WD mass, the shorter the predicted post-CE orbital periods.
Interestingly, post-CE binaries hosting WDs with masses lower than ${\sim0.5}$~\Msun~are virtually not predicted at all when CE evolution is highly inefficient.
Such post-CE binaries would need to form when the WD progenitor is on the FGB, which means that in most cases the onset of the CE evolution occurs at a relatively short orbital separation.
Therefore, if $\alpha$ is low, there is not enough orbital energy available to successfully unbind the CE, which results in a merger.

On the other hand, for highly efficient CE evolution (${\alpha\gtrsim0.9}$), post-CE orbital periods can be as long as a few hundred days for low-mass carbon-oxygen WDs, and one thousand days for oxygen-neon WDs.
The post-CE binaries with the longest orbital periods typically descend from pre-CE binaries with orbital periods longer than ${10^4}$~d, having TP-AGB donors with large radii (${\gtrsim1400}$~\Rsun) and tiny envelopes (${\lesssim0.5}$~\Msun).

Interestingly, many close post-CE binaries (orbital period shorter than a few days) host initially helium stars that later evolve into WDs \citep[see][for more details on how helium stars are defined and treated in the BSE code]{Hurley_2000}.
This happens when the CE evolution takes place with an early-AGB donor, which leads to the formation of a helium giant star as unburnt helium remains within the hydrogen-exhausted core causing shell helium burning through which the carbon core grows \citep[see also][]{Li_2024}.
In some cases, there is an additional episode of mass transfer if the helium star fills its Roche lobe before becoming a WD, and whether it is dynamically stable or not depends on the binary properties.
In general, the lowest predicted WD mass slightly decreases as $\alpha$ increases.

We further predict an accumulation of systems at shorter orbital periods (between ${\sim1-10}$~d) for WDs more massive than ${\sim0.9}$~\Msun; this accumulation is always present regardless of the value of  $\alpha$, but   becomes more pronounced when $\alpha$ increases.
This accumulation is a consequence of the stellar type of the WD progenitor at the onset of the CE evolution.
For those systems, CE evolution occurred when the WD progenitor was still ascending the AGB or had just become a TP-AGB star.
For all systems with longer orbital periods (i.e. above these accumulations in Fig.\ref{Fig:MwdvsPorb}), the onset of the CE evolution occurred when the donor was already an evolved TP-AGB star, with a less massive and less bound envelope.

There are two other   features in Fig.~\ref{Fig:MwdvsPorb}, one of which is very clear, corresponding to overabundances of systems with orbital periods ${\lesssim5}$~d and with WDs of masses ${\sim0.52-0.56}$ and ${\sim0.75-0.80}$~\Msun.
These two clumps of systems originate from zero-age binaries in which the WD progenitor has a mass of ${\sim2}$ and ${\sim3-4}$~\Msun, respectively.
The onset of the CE evolution in these cases takes place when the WD progenitor is an early AGB star and for this reason the resulting post-CE binary hosts initially a helium star.
This helium star then later quickly evolves into a WD.
These two overabundances are likely a consequence of the correlated distributions we adopted for the zero-age main-sequence binaries, that is, those derived from observations by \citet{MD17}.

When compared to the observational sample of long-period post-CE binaries hosting oxygen-neon WDs and \afgk~main-sequence stars, we can conclude that their properties can be explained for any CE efficiency ${\alpha\gtrsim0.2}$.
For efficiencies lower than that, the predicted maximum post-CE orbital period is typically shorter than those of observed systems.
Given that the location of the observed systems is above the accumulation of the post-CE binaries with massive WDs and short orbital periods we mentioned earlier, these systems must descend from systems where the CE evolution started when the donor star was an evolved TP-AGB star. 
We show in Sect.~\ref{FormChannel} that this is indeed the case.

Although our focus here is on post-CE binaries containing oxygen-neon WDs, we note that our simulations predict many long-period systems with carbon-oxygen WDs, which is consistent with observations.
For instance, the self-lensing post-CE binary KOI~3278 harbours a carbon-oxygen WD with a G-type main-sequence star and has an orbital period of ${\approx88}$~d \citep{KruseAgol_2014,Yahalomi_2019}.
More recently, \citet{Garbutt_2024} found a sizeable sample of long-period post-CE binaries hosting WDs with masses ${\sim0.6-0.9}$~\Msun.
The evolutionary history of these systems will be addressed in detail in a subsequent paper.

We can now turn to the discussion of the properties of the detached post-CE binaries in the plane main-sequence mass versus orbital period (Fig.~\ref{Fig:M2vsPorb}).
All known long-period post-CE binaries with oxygen-neon WDs and \afgk~main-sequence stars can be explained for any value of the CE efficiency ${\alpha\gtrsim0.2}$, similarly to what we concluded while analysing Fig.~\ref{Fig:MwdvsPorb}.

The predicted detached post-CE binaries are mainly found in two groups.
The first group corresponds to main-sequence stars with masses ${\gtrsim1.25}$~\Msun, which is due to the fact that magnetic braking is not expected to be acting in these stars.
This means that their orbital periods are less likely to change as the only source of orbital angular momentum loss is the emission of gravitational waves, which only affects binaries with sufficiently short orbital periods.
The second group is made up of systems with sufficiently long orbital periods (${\gtrsim5}$~d) so that regardless of whether magnetic braking is acting or not in these binaries, their long orbital periods prevent them from being affected by any source of orbital angular momentum loss.
In addition to these two main groups, the rest of the systems host main-sequence stars of masses ${\lesssim1.25}$~\Msun~and have orbital periods ${\lesssim5}$~d.
For them, magnetic braking together with gravitational wave radiation more easily converts them into semi-detached binaries, known as cataclysmic variables.
As a result, this region in this plane is less populated.

\subsection{Formation pathways}
\label{FormChannel}

We have just shown in Sect.~\ref{PCEBprop} that post-CE binaries hosting oxygen-neon WDs can have orbital periods from several tens of days (for ${\alpha\sim0.1}$)   to one thousand days (for ${\alpha\sim1.0}$), without including energy sources other than orbital, gravitational, and thermal in the CE energy budget.
We now turn our discussion to the specific formation pathways of the six known post-CE binaries we have been discussing.
We were able to successfully find very decent models for each of the six observed post-CE binaries listed in Table~\ref{TableOBS}.
We include in the Appendix~\ref{APTables} examples of formation pathways for each  of them (Tables~\ref{Tab:FormationChannel:IKPeg}--\ref{Tab:FormationChannel:J0107}).

We searched for best-fitting models using the BSE code by carrying out pre-CE and CE evolution adopting the assumptions described in Sect.~\ref{BSE} and assuming that the zero-age main-sequence binary orbit was circular.
We set the CE efficiency to ${\alpha=0.3}$, which is consistent with the increasing evidence that short-period post-CE binary progenitors experience strong orbital shrinkage during CE evolution \citep[e.g.][]{Zorotovic_2010,Toonen_2013,Camacho_2014,Cojocaru_2017,Belloni_2019,hernandezetal22-1, Zorotovic_2022,schrebak+fuller23-1}.
For each observed system, we ran a large grid of binary models varying the zero-age mass of the WD progenitor from $6$ to $8$~\Msun, in steps of $0.01$~\Msun~and the zero-age orbital period from $10^3$ to $10^4$~d, in steps of $5$~d.
Finally, the zero-age mass of the companion was chosen to be slightly lower than the observed values as it increases during binary evolution due to wind accretion.
After finding within the grid the model closest to the observed one, we made the grid finer around this model to finally obtain a best-fitting model.
We note that the models provided in Tables~\ref{Tab:FormationChannel:IKPeg}--\ref{Tab:FormationChannel:J0107} are not the only best-fitting models.
Instead, each  model belongs to a family of solutions able to explain the observed system; this family   arises from varying the other assumptions such as zero-age eccentricity, stability of mass transfer, wind accretion efficiency and model, and metallicity, among others.

Even so, these families of models share the same fundamental features, which characterize the formation pathways of long-period post-CE binaries hosting oxygen-neon WDs and \afgk~main-sequence stars.
As these systems host oxygen-neon WDs, for solar metallicity and the assumptions within the BSE code for stellar evolution, the initial mass of their progenitors must have been ${\gtrsim6}$~\Msun.
Such stars lose only a negligible amount of mass before becoming a TP-AGB star.
This means that they start the TP-AGB phase with a very massive and strongly bound envelope.
If they fill their Roche lobe with a large envelope mass fraction, the resulting post-CE orbital period will   necessarily be much shorter than those observed among these six known systems.

In all cases, to reproduce the observed long orbital periods, two main conditions at the onset of the CE evolution are required.
First, the WD progenitor has to fill its Roche lobe as it is a highly evolved TP-AGB star (i.e. when more than 50\% of its mass has already been lost through stellar winds), resulting in an envelope-structure parameter of ${\sim1.0-1.2}$.
Second, the orbital period has to be several thousand days  because highly evolved TP-AGB stars can develop more loosely bound envelopes, and fill their Roche lobes at sufficiently long orbital periods so that the fraction of the available orbital energy that is required to eject the envelope does not lead to a strong orbital shrinkage.
Otherwise, the resulting post-CE binary would have a much shorter orbital period.

Most importantly, unlike what has been claimed for a long time, no extra energy is required to explain the well-known system IK~Peg and the six recently discovered systems.
Therefore, our results provide further support for inefficient CE evolution as the most common formation channel of post-CE binaries hosting WDs, regardless of whether they have short or long orbital periods.

\section{Discussion}
\label{Discussion}

\subsection{Mass transfer from red giants in the BSE code}
\label{BSEvsStability}

The BSE code is based on analytic formulae that approximate the evolution of stars \citep{Hurley_2000} and are, in general, reasonably accurate for low- and intermediate-mass stars.
In addition, for the TP-AGB phase, BSE does not model the thermal pulses individually and only takes into account the third dredge-ups, which correspond to the thermally pulsing behaviour on the long-term evolution.
Therefore, future modelling efforts with codes that can resolve the thermal pulses will be useful to further test the results achieved in this work.

Regarding the stability of mass transfer from TP-AGB stars, the BSE code adopts a criterion based on models of condensed polytropes \citep{Hjellming_1987}, for which fully conservative mass transfer is assumed.
However, \citet{Ge_2020} showed that these assumptions lead to red giants that tend to be less dynamically stable in comparison with more accurate models.

To further progress with the study of post-CE binaries descending from TP-AGB stars, it is crucial to properly establish consistent criteria for the stability of mass transfer from these stars.
A promising route towards a better prescription for the stability of mass transfer from TP-AGB donors could be a joint criterion taking into account dynamical \citep{Ge_2020} and thermal timescale mass transfer \citep{Ge_2020b}.
However, even incorporating   a new prescription into BSE might not cover the full complexity of mass transfer from 
TP-AGB stars. Until full binary population synthesis with codes as precise as MESA are feasible, it might be impossible to overcome all limitations.

\subsection{Impact of the adopted envelope-structure parameter}
\label{BSEvsLambda}

One might wonder whether the BSE code inconsistently estimates $\lambda$, and that this is perhaps the reason why we managed to reproduce the known long-period post-CE binaries with oxygen-neon WDs and \afgk~main-sequence stars.
The prescription adopted in BSE to compute $\lambda$ is based on the detailed calculations by \citet{Dewi_2000}.
As already pointed out by \citet{Davis_2010}, the tabulated values provided by \citet{Dewi_2000} only cover stars of masses ${\geq3}$~\Msun~with radii up to ${\sim400-600}$~\Rsun.
This means that these authors halted their calculations before the stars entered the TP-AGB phase.

\begin{figure}
\begin{center}
\includegraphics[width=0.99\linewidth]{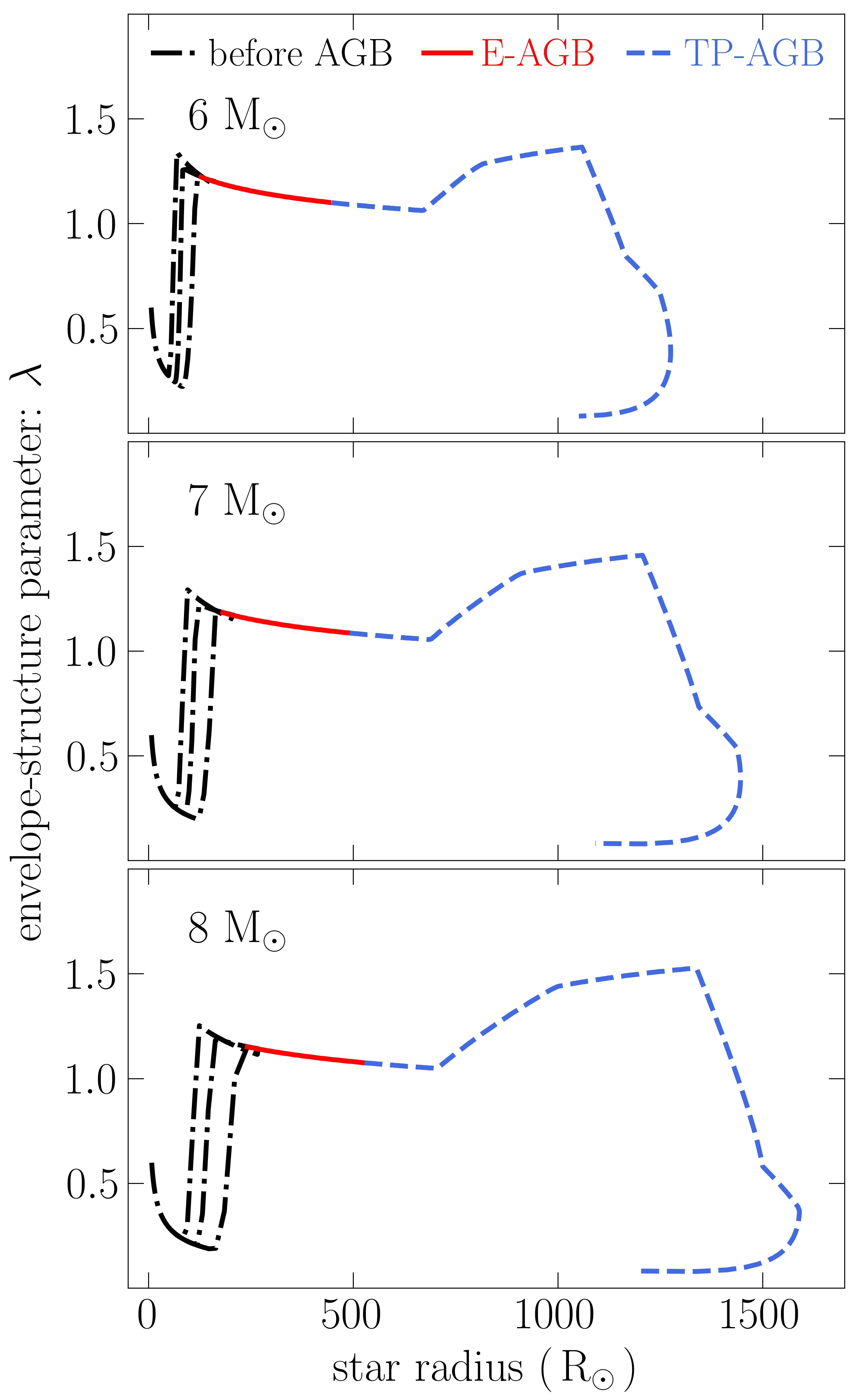}
\end{center}
\caption{Evolution of the envelope structure parameter $\lambda$ with the star radius for different initial masses that are progenitors of the most massive WDs, calculated with the BSE code including only gravitational and thermal energy, for the before AGB (black), early AGB (red), and TP-AGB phases (blue). The values of $\lambda$ estimated by the BSE code for TP-AGB stars smoothly increase as the star evolves, but are comparable with the values when the star is on the early AGB. In addition, for all these masses, ${\lambda\sim1.0-1.5}$ during the TP-AGB phase. Only when the star is very close to becoming a WD does $\lambda$ drop, which is a consequence of  a very large core mass fraction at the end of their lives.}
\label{Fig:BSElambda}
\end{figure}

The fitting scheme adopted in BSE, which is similar to the formulas provided by \citet{Claeys_2014}, results in values of $\lambda$ that smoothly increases for the TP-AGB phase as the star evolves, and are similar to the values just before the star becomes a TP-AGB star.
This is illustrated in Fig.~\ref{Fig:BSElambda}, where we show the dependence of $\lambda$ on the stellar radius for stars with initial masses of $6$, $7,$ and $8$~\Msun, which are the progenitors of the most massive WDs (according to BSE).
In all cases the estimated value for $\lambda$ during most of the TP-AGB phase is in the range ${\sim1.0-1.5}$, and we found no problem to reproduce the observed systems with these values (Tables~\ref{Tab:FormationChannel:IKPeg}--\ref{Tab:FormationChannel:J0107}).
Only when the mass of the stellar envelope is very low, that is, when the star is close to becoming a WD, the value of $\lambda$ drops.

To illustrate that our results are independent of the assumed scheme for computing the envelope-structure parameter, we repeated here the search for examples of formation pathways leading to all six observed systems, but now fixing the value of $\lambda$ to $0.25$, which is a value consistent with an FGB star and, in turn, much smaller than expected for a TP-AGB star.
We then inspected whether we could find   a reasonable model for each system, and under what conditions (e.g. for what values of the CE efficiency $\alpha$).
All systems can be reproduced assuming a CE efficiency of at least ${\sim0.2-0.3}$.
We can therefore conclude that our results remain valid, even if we assume that the BSE code overestimates $\lambda$, which does not seem to be the case as the values we get here are consistent with others \citep[e.g.][]{Xu_2010,Loveridge_2011,Ablimit_2016}.
In addition, this exercise also indicates that all observed systems could be explained even when only orbital and gravitational energies are included in the CE evolution energy budget, that is, with ${\lambda\sim0.5-1.0}$.

\begin{figure}
  \begin{center}
    \includegraphics[width=0.99\linewidth]{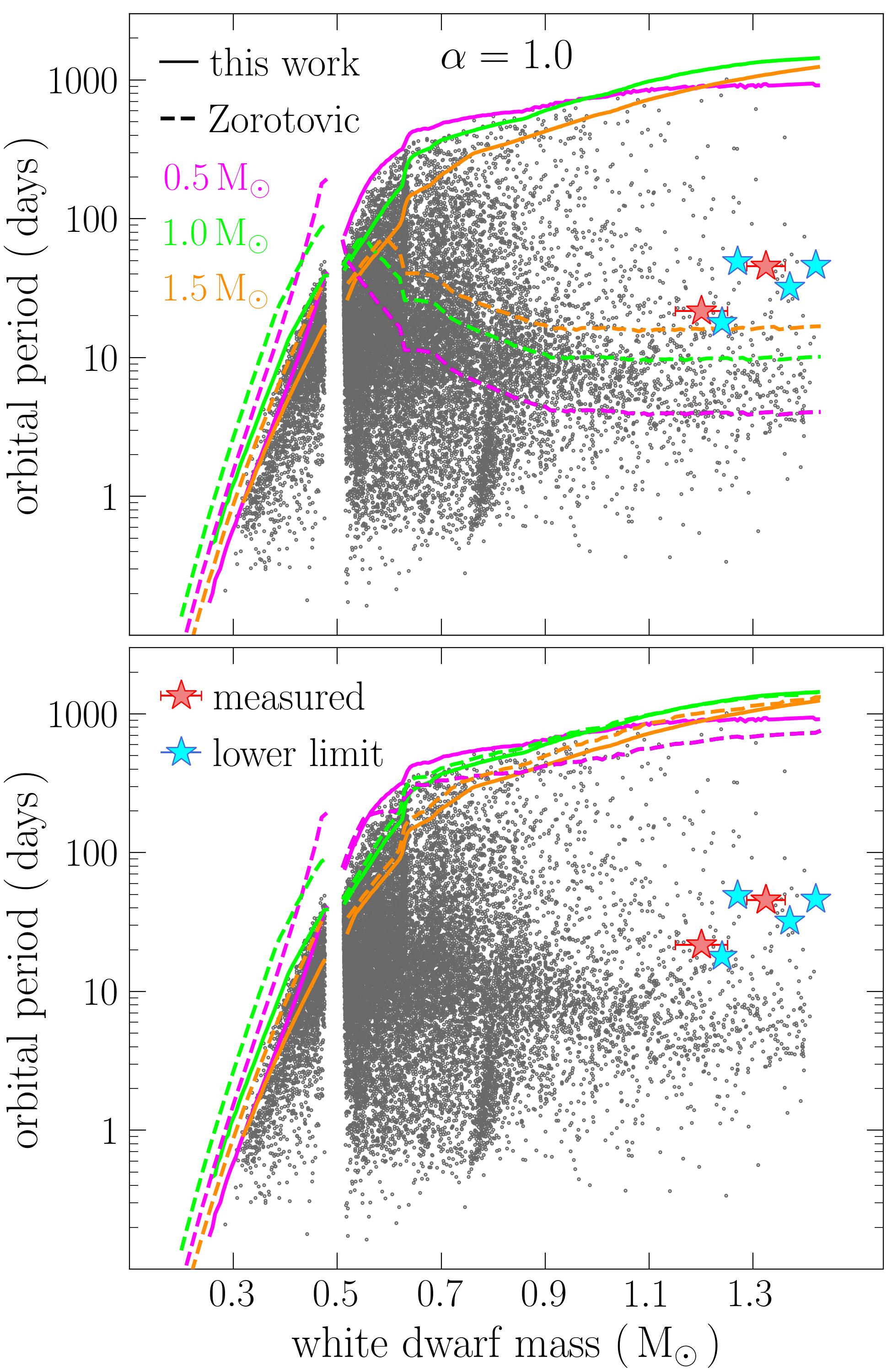}
  \end{center}
  \caption{Maximum orbital period just after CE evolution as a function of WD mass computed in this work (solid lines) and using the algorithm based on a grid for possible progenitors of the WDs from Zorotovic (dashed lines) plotted against the outcome of our post-CE binary population synthesis assuming ${\alpha=1}$ and no other sources of energy than gravitational and thermal. The colours of the lines and markers are the same as those in Fig.~\ref{Fig:MwdvsPorb}. In the top panel the dashed lines are calculated restricting the progenitor's luminosity to be lower than the peak luminosity of the first thermal pulse (as in e.g. \citealt{Rebassa-Mansergas2012}), while in the bottom panel this restriction is not used, allowing for progenitors on the TP-AGB phase. It is evident that the approach employed by Zorotovic leads to much shorter maximum orbital periods for systems with massive WDs when excluding progenitors on the TP-AGB in comparison to our method. This explains why it was not possible to reproduce the long-period post-CE binaries with oxygen-neon WDs using that method. By removing the luminosity restriction and allowing progenitors in the TP-AGB phase, the results obtained with the Zorotovic's algorithm are very similar to our results.}
  \label{Fig:BSEvsMonica}
\end{figure}

\subsection{Comparison with previous works}
\label{WhyTheyFailed}

The discussion concerning possible contributions of the ionization or recombination energy to the ejection of planetary nebulae during single-star evolution started in the 1960s \citep{Lucy_1967,Roxburgh_1967,Paczynski_1968}.
In the 1990s \citet{Han_1994} and \citet{Han_1995} further investigated the importance of the ionization energy of hydrogen and helium to the internal energy of the envelopes not only of single stars, but also in stars undergoing CE evolution.
Whether recombination energy can have a significant impact on the CE ejection  has since then been an active topic of research \citep[e.g.][and references therein]{Webbink_2008,Ivanova_REVIEW,Ropke_2023}.


To the best of our knowledge, \citet{Davis_2010} was the first to claim that extra energy is required to explain the properties of IK~Peg, which was the only long-period post-CE binary hosting an oxygen-neon WD known before Gaia DR3.
They carried out post-CE population synthesis adopting a similar scheme to ours to calculate their envelope-structure parameters, that is, based on the detailed calculations performed by \citet{Dewi_2000}, although these authors interpolated the tabulated values provided by \citet{Dewi_2000} and extended their grid with the same code.
They also used a set of ${5.4\times10^5}$ zero-age main-sequence binaries, which is a number comparable to ours.

In our post-CE binary population synthesis, whose results are illustrated in Figs.~\ref{Fig:MwdvsPorb} and \ref{Fig:M2vsPorb}, we predict the existence of long-period post-CE binaries similar to the six systems we address here.
It is not clear to us why \citet{Davis_2010} failed to predict the existence of such systems.
The disagreement might be related to ambiguity in several terms used by \citet{Davis_2010}.
For instance, it is difficult to understand what they call `thermal energy', `internal energy', `recombination energy', `ionization energy', and `extra energy'.
Because of this problem, discussing in greater detail the differences of our results to those obtained by \citet{Davis_2010} appears to be a rather futile exercise.


In the same year, \citet{Zorotovic_2010} also claimed that IK\,Peg needed extra energy sources, based on a similar approach to that used by \citet{Davis_2010}.
However, as stated in \citet{Zorotovic_2010}, they required the progenitors to have a luminosity lower than the peak luminosity of the first thermal pulse.
Therefore, potential progenitors during the subsequent pulses were not considered.
In the top panel of Fig.~\ref{Fig:BSEvsMonica} we show a comparison between our maximum post-CE orbital periods and those from the algorithm developed by Zorotovic (first used and explained in \citealt{Rebassa-Mansergas2012}, their Fig. 6).
We can see in the figure that they differ quite significantly for post-CE binaries hosting carbon-oxygen WDs and oxygen-neon WDs.
The limits derived by Zorotovic correspond to shorter orbital periods (much shorter than the observed systems discussed in this work). This is not surprising as we   show  in the previous sections that evolved TP-AGB  donors are required to reproduce the observed systems.

However, eliminating the condition that the progenitor's luminosity should be lower than the peak luminosity of the first thermal pulse from the algorithm developed by Zorotovic, the maximum post-CE orbital periods we calculated here with the BSE code and those computed with   Zorotovic's method agree reasonably well, as illustrated in the bottom panel of Fig.~\ref{Fig:BSEvsMonica}.
The small differences are likely associated with accretion by the secondary prior to the CE phase, a factor that cannot be accounted for in   Zorotovic's algorithm as it is based on a grid of progenitors generated with the SSE code from \citet{Hurley_2000}.
This indicates that assuming a grid of progenitors based on single stellar evolution, as done by Zorotovic, leads to decent results, provided all potential progenitors of all types of WDs are included.


\citet{Yamaguchi_2024} claims that highly efficient CE evolution, with the inclusion of a significant fraction of recombination energy, is needed to explain the five long-period binaries they discovered from {\it Gaia} DR3 containing oxygen-neon WDs  (and which we addressed in this work).
One clear aspect of the formation pathways we found here is that the WD progenitors have to be highly evolved TP-AGB stars at the onset of the CE evolution.
This means that the red giant had enough time to lose a significant fraction of its mass (${\gtrsim50}$\%) before filling its Roche lobe.
\citet{Yamaguchi_2024} used the MESA code to evolve a $7$~\Msun~pre-main-sequence star up to the AGB.
In other words, these authors did not follow the evolution of the star during the TP-AGB phase as they stopped their simulation before carbon ignition when the star was not as big as it could get and had lost virtually no mass through stellar winds.
This is most likely the reason why \citet{Yamaguchi_2024} needed to include recombination energy and to assume highly efficient CE evolution in their calculations to reproduce the five systems and IK~Peg.

\section{Conclusions}\label{Conclusion}

We carried out post-CE binary population model with the BSE code, with the aim of determining whether energy sources other than orbital, gravitational, and thermal should be included in the CE evolution energy budget to reproduce the six known long-period post-CE binaries with oxygen-neon WDs and \afgk~main-sequence stars.
We found that all six systems can be reproduced reasonably well with inefficient CE evolution without invoking extra energy.
This was achieved by allowing the WD progenitors to become highly evolved TP-AGB stars before filling its Roche lobe, at the onset of the CE evolution.
We also found that post-CE binaries can have orbital periods as long as one thousand days when all available orbital energy is used to unbind the CE.
Our results provide further evidence for a common origin of post-CE binaries with WD primaries; in other words,  it seems all observed systems are consistent with having formed through  inefficient CE evolution, irrespective of their post-CE orbital period.

\begin{acknowledgements}

We would like to thank the referee Hongwei Ge for his comments and suggestions that helped to improve this manuscript.
We thank the Kavli Institute for Theoretical Physics (KITP) for hosting the program ``White Dwarfs as Probes of the Evolution of Planets, Stars, the Milky Way and the Expanding Universe''.
This research was supported in part by the National Science Foundation under Grant No. NSF PHY-1748958.
This research was partially supported by the Munich Institute for Astro-, Particle and BioPhysics (MIAPbP) which is funded by the Deutsche Forschungsgemeinschaft (DFG, German Research Foundation) under Germany's Excellence Strategy -- EXC--2094 -- 390783311.
DB acknowledges financial support from {FONDECYT} grant number {3220167}.
MZ was supported by {FONDECYT} grant number {1221059}.
MRS was supported by {FONDECYT} grant number {1221059} and ANID, – Millennium Science Initiative Program – NCN19\_171.

\end{acknowledgements}

%
%

\bibliographystyle{aa} 
\bibliography{references} 


\begin{appendix}

\section{Examples of formation pathways for the known long-period post-CE binaries with oxygen-neon WDs and AFGK-type main-sequence stars}
\label{APTables}

\begin{table*}
\centering
\caption{Evolution of a zero-age main-sequence binary towards IK~Peg. For the pre-CE evolution, CE evolution, and post-CE evolution, we used the BSE code. We adopted the assumptions described in Sect.~\ref{BSE}, a CE efficiency of ${\alpha=0.3}$ and a circular  zero-age main-sequence binary orbit. The terms $M_1$ and $M_2$ and Type$_1$ and Type$_2$ are the masses and stellar types$^{a}$ of the progenitors of the WD and companion, respectively, and $R_1$ is the radius of the WD progenitor. $P_{\rm orb}$ is the orbital period. The  last column corresponds to the event occurring in the binary at the time given in the first column.}
\label{Tab:FormationChannel:IKPeg}
\setlength\tabcolsep{9pt} 
\renewcommand{\arraystretch}{1.25} 
\begin{tabular}{r c c r r c r l}
\hline
\noalign{\smallskip}
 Time  &   $M_1$    &   $M_2$    &   $R_1$    & Type$_1$  & Type$_2$ & Orbital Period & Event\\
 (Myr) & (M$_\odot$)&(M$_\odot$) &(R$_\odot$) &          &  (days)        &      \\
\hline
\noalign{\smallskip}
     0.0000  &  6.510  &  1.600  &    3.066   & MS     & MS  &   6108.700  &  zero-age MS binary \\
    57.3386  &  6.481  &  1.600  &    7.148   & SG     & MS  &   6152.441  &  change in primary type \\
    57.5368  &  6.481  &  1.600  &   82.463   & FGB    & MS  &   6153.224  &  change in primary type \\
    57.6120  &  6.479  &  1.600  &  187.667   & CHeB   & MS  &   6155.216  &  change in primary type \\
    64.9580  &  6.362  &  1.600  &  154.565   & E-AGB  & MS  &   6336.515  &  change in primary type \\
    65.3065  &  6.313  &  1.600  &  470.502   & TP-AGB & MS  &   6398.505  &  change in primary type \\
    65.7503  &  3.527  &  1.706  & 1055.410   & TP-AGB & MS  &   5883.777  &  begin RLOF (primary is the donor) \\
    65.7503  &  3.527  &  1.706  & 1055.410   & TP-AGB & MS  &   5883.777  &  CE evolution $\left({\lambda=1.2066}\right)$\\
    65.7503  &  1.200  &  1.706  &    0.006   & WD     & MS  &     21.717  &  end RLOF \\
\noalign{\smallskip}
\hline
\end{tabular}
\tablefoot{\tablefoottext{a}{Abbreviations:
MS~(main-sequence~star),
SG~(subgiant~star),
FGB~(first giant branch star),
CHeB~(core helium burning),
E-AGB~(early asymptotic giant branch star),
TP-AGB~(thermally pulsing~asymptotic~giant~branch~star),
WD~(white~dwarf),
RLOF~(Roche-lobe~overflow),
CE~(common~envelope).}}
\end{table*}

\begin{table*}
\centering
\caption{Same as Table~\ref{Tab:FormationChannel:IKPeg}, but for J2117$+$0332.}
\label{Tab:FormationChannel:J2117}
\setlength\tabcolsep{9pt} 
\renewcommand{\arraystretch}{1.25} 
\begin{tabular}{r c c r r c r l}
\hline
\noalign{\smallskip}
 Time  &   $M_1$    &   $M_2$    &   $R_1$    & Type$_1$  & Type$_2$ & Orbital Period & Event\\
 (Myr) & (M$_\odot$)&(M$_\odot$) &(R$_\odot$) &           &          &  (days)        &      \\
\hline
\noalign{\smallskip}
     0.0000  &  6.810  &  1.050  &    3.147   & MS     & MS  &   6643.650  &  zero-age MS binary \\
    52.0696  &  6.774  &  1.050  &    7.341   & SG     & MS  &   6704.915  &  change in primary type \\
    52.2444  &  6.774  &  1.050  &   90.457   & FGB    & MS  &   6705.872  &  change in primary type \\
    52.3055  &  6.772  &  1.050  &  203.319   & CHeB   & MS  &   6708.251  &  change in primary type \\
    58.8623  &  6.631  &  1.050  &  170.864   & E-AGB  & MS  &   6955.123  &  change in primary type \\
    59.1647  &  6.580  &  1.050  &  483.530   & TP-AGB & MS  &   7039.943  &  change in primary type \\
    59.6036  &  3.369  &  1.109  & 1123.476   & TP-AGB & MS  &   6247.009  &  begin RLOF (primary is the donor) \\
    59.6036  &  3.369  &  1.109  & 1123.476   & TP-AGB & MS  &   6247.009  &  CE evolution $\left({\lambda=1.1998}\right)$\\
    59.6036  &  1.244  &  1.109  &    0.005   & WD     & MS  &     17.925  &  end RLOF \\
\noalign{\smallskip}
\hline
\end{tabular}
\end{table*}

\begin{table*}
\centering
\caption{Same as Table~\ref{Tab:FormationChannel:IKPeg}, but for J1111$+$5515.}
\label{Tab:FormationChannel:J1111}
\setlength\tabcolsep{9pt} 
\renewcommand{\arraystretch}{1.25} 
\begin{tabular}{r c c r r c r l}
\hline
\noalign{\smallskip}
 Time  &   $M_1$    &   $M_2$    &   $R_1$    & Type$_1$  & Type$_2$ & Orbital Period & Event\\
 (Myr) & (M$_\odot$)&(M$_\odot$) &(R$_\odot$) &           &          &  (days)        &      \\
\hline
\noalign{\smallskip}
     0.0000  &  7.630  &  1.075  &    3.362   & MS     & MS  &   6915.230  &  zero-age MS binary \\
    41.1449  &  7.568  &  1.075  &    7.856   & SG     & MS  &   7015.739  &  change in primary type \\
    41.2735  &  7.567  &  1.075  &  113.531   & FGB    & MS  &   7017.309  &  change in primary type \\
    41.3097  &  7.565  &  1.075  &  247.530   & CHeB   & MS  &   7019.944  &  change in primary type \\
    46.3150  &  7.355  &  1.075  &  218.816   & E-AGB  & MS  &   7372.224  &  change in primary type \\
    46.5246  &  7.298  &  1.075  &  515.337   & TP-AGB & MS  &   7464.184  &  change in primary type \\
    46.9537  &  3.232  &  1.149  & 1273.334   & TP-AGB & MS  &   7788.400  &  begin RLOF (primary is the donor) \\
    46.9537  &  3.232  &  1.149  & 1273.334   & TP-AGB & MS  &   7788.400  &  CE evolution $\left({\lambda=1.1877}\right)$\\
    46.9537  &  1.367  &  1.149  &    0.003   & WD     & MS  &     32.145  &  end RLOF \\
\noalign{\smallskip}
\hline
\end{tabular}
\end{table*}

\begin{table*}
\centering
\caption{Same as Table~\ref{Tab:FormationChannel:IKPeg}, but for J1314$+$3818.}
\label{Tab:FormationChannel:J1314}
\setlength\tabcolsep{9pt} 
\renewcommand{\arraystretch}{1.25} 
\begin{tabular}{r c c r r c r l}
\hline
\noalign{\smallskip}
 Time  &   $M_1$    &   $M_2$    &   $R_1$    & Type$_1$  & Type$_2$ & Orbital Period & Event\\
 (Myr) & (M$_\odot$)&(M$_\odot$) &(R$_\odot$) &           &          &  (days)        &      \\
\hline
\noalign{\smallskip}
     0.0000  &  7.340  &  0.660  &    3.287   & MS     & MS  &   6662.420  &  zero-age MS binary \\
    44.5261  &  7.288  &  0.660  &    7.676   & SG     & MS  &   6749.692  &  change in primary type \\
    44.6686  &  7.287  &  0.660  &  105.182   & FGB    & MS  &   6750.985  &  change in primary type \\
    44.7120  &  7.286  &  0.660  &  231.695   & CHeB   & MS  &   6753.604  &  change in primary type \\
    50.1852  &  7.099  &  0.660  &  201.359   & E-AGB  & MS  &   7082.557  &  change in primary type \\
    50.4230  &  7.043  &  0.660  &  504.838   & TP-AGB & MS  &   7178.127  &  change in primary type \\
    50.8624  &  2.443  &  0.708  & 1336.698   & TP-AGB & MS  &   9340.517  &  begin RLOF (primary is the donor) \\
    50.8624  &  2.443  &  0.708  & 1336.698   & TP-AGB & MS  &   9340.517  &  CE evolution $\left({\lambda=1.0128}\right)$\\
    50.8624  &  1.324  &  0.708  &    0.004   & WD     & MS  &     45.470  &  end RLOF \\
\noalign{\smallskip}
\hline
\end{tabular}
\end{table*}

\begin{table*}
\centering
\caption{Same as Table~\ref{Tab:FormationChannel:IKPeg}, but for J2034$-$5037.}
\label{Tab:FormationChannel:J2034}
\setlength\tabcolsep{9pt} 
\renewcommand{\arraystretch}{1.25} 
\begin{tabular}{r c c r r c r l}
\hline
\noalign{\smallskip}
 Time  &   $M_1$    &   $M_2$    &   $R_1$    & Type$_1$  & Type$_2$ & Orbital Period & Event\\
 (Myr) & (M$_\odot$)&(M$_\odot$) &(R$_\odot$) &           &          &  (days)        &      \\
\hline
\noalign{\smallskip}
     0.0000  &  7.970  &  0.890  &    3.449   & MS     & MS  &   6922.540  &  zero-age MS binary \\
    37.7079  &  7.893  &  0.890  &    8.065   & SG     & MS  &   7044.587  &  change in primary type \\
    37.8226  &  7.892  &  0.890  &  123.551   & FGB    & MS  &   7046.622  &  change in primary type \\
    37.8522  &  7.890  &  0.890  &  266.306   & CHeB   & MS  &   7049.384  &  change in primary type \\
    42.3932  &  7.652  &  0.890  &  239.803   & E-AGB  & MS  &   7446.571  &  change in primary type \\
    42.5744  &  7.593  &  0.890  &  526.636   & TP-AGB & MS  &   7542.258  &  change in primary type \\
    43.0050  &  2.880  &  0.957  & 1375.186   & TP-AGB & MS  &   9164.345  &  begin RLOF (primary is the donor) \\
    43.0050  &  2.880  &  0.957  & 1375.186   & TP-AGB & MS  &   9164.345  &  CE evolution $\left({\lambda=1.1127}\right)$\\
    43.0050  &  1.419  &  0.957  &    0.002   & WD     & MS  &     46.106  &  end RLOF \\
\noalign{\smallskip}
\hline
\end{tabular}
\end{table*}

\begin{table*}
\centering
\caption{Same as Table~\ref{Tab:FormationChannel:IKPeg}, but for J0107$-$2827.}
\label{Tab:FormationChannel:J0107}
\setlength\tabcolsep{9pt} 
\renewcommand{\arraystretch}{1.25} 
\begin{tabular}{r c c r r c r l}
\hline
\noalign{\smallskip}
 Time  &   $M_1$    &   $M_2$    &   $R_1$    & Type$_1$  & Type$_2$ & Orbital Period & Event\\
 (Myr) & (M$_\odot$)&(M$_\odot$) &(R$_\odot$) &           &          &  (days)        &      \\
\hline
\noalign{\smallskip}
     0.0000  &  6.990  &  0.900  &    3.195   & MS     & MS  &   6761.780  &  zero-age MS binary \\
    49.2816  &  6.949  &  0.900  &    7.455   & SG     & MS  &   6832.432  &  change in primary type \\
    49.4442  &  6.949  &  0.900  &   95.376   & FGB    & MS  &   6833.454  &  change in primary type \\
    49.4985  &  6.947  &  0.900  &  212.863   & CHeB   & MS  &   6835.959  &  change in primary type \\
    55.6484  &  6.791  &  0.900  &  180.994   & E-AGB  & MS  &   7115.337  &  change in primary type \\
    55.9267  &  6.738  &  0.900  &  491.036   & TP-AGB & MS  &   7206.304  &  change in primary type \\
    56.3706  &  2.552  &  0.975  & 1261.822   & TP-AGB & MS  &   8739.329  &  begin RLOF (primary is the donor) \\
    56.3706  &  2.552  &  0.975  & 1261.822   & TP-AGB & MS  &   8739.329  &  CE evolution $\left({\lambda=1.0566}\right)$\\
    56.3706  &  1.272  &  0.975  &    0.005   & WD     & MS  &     49.015  &  end RLOF \\
\noalign{\smallskip}
\hline
\end{tabular}
\end{table*}

\end{appendix}

\end{document}